\hfuzz 2pt
\font\titlefont=cmbx10 scaled\magstep1
\font\bigscript=cmbxsl10 scaled\magstep1
\font\script=cmbxsl10
\magnification=\magstep1
\null
\vskip 1.5cm
\centerline{\titlefont {\bigscript CPT}, DISSIPATION, AND ALL THAT}
\vskip 2.5cm
\centerline{\bf F. Benatti}
\smallskip
\centerline{Dipartimento di Fisica Teorica, Universit\`a di Trieste}
\centerline{Strada Costiera 11, 34014 Trieste, Italy}
\centerline{and}
\centerline{Istituto Nazionale di Fisica Nucleare, Sezione di
Trieste}
\vskip 1cm
\centerline{\bf R. Floreanini}
\smallskip
\centerline{Istituto Nazionale di Fisica Nucleare, Sezione di
Trieste}
\centerline{Dipartimento di Fisica Teorica, Universit\`a di Trieste}
\centerline{Strada Costiera 11, 34014 Trieste, Italy}
\vskip 2cm
\centerline{\bf Abstract}
\smallskip
\midinsert
\narrower\narrower\noindent
A phenomenological paradigm for the study of $CPT$-violating
effects in the neutral kaon system is presented. Besides the familiar
direct and indirect breakings, it encodes possible phenomena leading
to irreversibility and dissipation, that could originate from quantum
effects at Planck's scale. These new effects can be experimentally
probed with great accuracy, in particular at $\phi$-factories.
\endinsert
\bigskip
\vfil\eject

\noindent
{\bf 1. INTRODUCTION}

\bigskip
\noindent
One of the most striking properties that can be derived from the general
structure of quantum field theory is the invariance under the combination
of charge conjugation ($C$), parity reflection ($P$), and time reversal
($T$) transformations. This result is based on a rigorous formulation
of the theory, that incorporates in form of assumptions
(the Wightman axioms) very general physical requirements, that any
sensible quantum field model should satisfy; these include:
existence of a (cyclic) vacuum state, positivity of energy (spectral
condition), covariance under the action of the restricted Poincar\'e
group.

A further condition is necessary in order to guarantee the validity
of Einstein's causality; it is usually called locality
(or microcausality) and essentially states that the field variables commute
(or anticommute) at space-like separated points. Actually, this
assumption can be substituted by the weak locality condition,
a much less restrictive requirement:
it states the commutativity in the two-point (more in general, $n$-point)
correlation functions for space-like separated points.
Together with the (Wightman) axioms mentioned before, this very mild
assumption is equivalent to the invariance under $CPT$ transformations:
this is the content of the $CPT$ theorem. (For a precise formulation and
discussions, see [1-3].)

In view of this result, $CPT$-invariance seems unavoidable in any sensible
model based on the relativistic quantum theory of fields. Since our present
description of elementary particle physics phenomena is based on the
standard model (and field extensions of it), the symmetry under
$CPT$ transformations appears automatic, at least from the theoretical point
of view.

This situation can be summarized by saying that $CPT$-invariance is
almost a ``kinematical'' property of relativistic quantum field theory.[4]
In view of this, any experimental observation of a violation of the
$CPT$-symmetry would clearly represent a strong evidence of unconventional
physics, and signal the need of a profound modification of the present
interpretation of high energy phenomena in terms of local,
relativistic fields.

General consequences of $CPT$-invariance are the equality of mass
and lifetime of particles and their relative antiparticles; and indeed,
most of the experimental effort in testing the symmetry under $CPT$
transformations has been devoted to check this equality for various
elementary particles.[5] Although some of the quoted bounds coming from
these tests look impressive (in particular, the one involving the
$K^0$-$\overline{K^0}$ mass difference), one should keep in mind
that they can be translated into direct bounds on possible $CPT$-violating
effects only with the help of additional considerations.[6-9]

In order to improve the present experimental limits on $CPT$-invariance,
some theoretical guidance is clearly needed; in particular, one should
provide indications on what phenomena should be looked at and consequently
on what kind of experiment should be carried out and at which accuracy.
In this respect, the best solution would be the construction of a microscopic
theory, valid at the fundamental level; this would then allow the calculation
from first principles of the phenomenological parameters, with predictions
about size and quality of the various signals, and further permit a direct
comparison of the results coming from different experiments.

Unfortunately, such a predictive theory does not yet exist; nevertheless,
various treatments of $CPT$-violating effects have been proposed in
the literature. Although inspired by more fundamental dynamics (strings,
branes), they are all phenomenological in nature. Despite their limited
predictive power, they can be of great help in analyzing data from present
and future experiments.

\vskip 2cm

\noindent
\line{{\bf 2. TOWARDS A UNIFIED PARADIGM FOR
{\script CPT }-VIOLATION}\hfill}
%\line{\phantom{\bf\indent 2. }{\bf OSCILLATIONS}\hfill}

\bigskip
\noindent
As already said, within standard quantum field theory there seems to be
little space for interactions violating $CPT$-symmetry. However,
there are technical assumptions in the $CPT$ theorem that, once released,
could open the possibility to $CPT$-violating effects.

One of these implicit conditions requires the field variables to have
a finite number of components, and therefore to transform according
to finite-dimensional representations under the action of the restricted
Lorentz group. Thus, infinite component field models can violate
$CPT$-invariance.[10]

The first example was presented in the early developments of quantum
field theory: indeed, Majorana wrote down a model of infinite-dimensional
wave equation without negative-energy solutions of non-negative square mass,
{\it i.e.} without antiparticles.[11] Since then, other more realistic theories
have been constructed$\,$%
\footnote{$^\dagger$}{The explicit construction of these models also
clarifies a point often misunderstood in the literature: $CPT$-invariance
and the spin-statistics connection are independent and separate
issues. For a recent discussion, see [12].}
and their quantization studied.[13, 14] Unfortunately,
these models either have an infinite mass degeneracy with respect to spin,
or violate locality. More precisely, one can show that (composite) systems
with an infinite degeneracy in the mass spectrum do not admit a simple
description in terms of local fields, even with infinite components;
therefore, the correct treatment of these models requires either
the weakening of the assumption of strict locality,
or the replacement of the field variables by some sort of extended
objects.[10]

This brings up another possible way to circumvent the assumptions of the $CPT$
theorem: construct field theories which present some sort of non-locality.
This possibility is not as exotic as it might appear at first:
after all, we are quite accustomed to the quantization of Yang-Mills
theories in non-local (Coulomb like) gauges; furthermore, a
rigorous treatment of quantum gauge theories might even require a reformulation
of the Wightman axioms. In any case, the non-local properties that these
theories contain are generally not enough to produce $CPT$-violating
effects.[3]

Nevertheless, non-local field models without $CPT$-symmetry have been
explicitly
constructed; they are Lorentz covariant and are quantized in the standard
way, but contain non-local field operators, due to an unconventional
interplay between charge conjugation and half-integer spin.[15]
Unfortunately, the kind of multiplets these theories describe does not seem
part of our phenomenological world.

As mentioned before in connection with infinite component fields,
the conclusions of the $CPT$ theorem can be avoided by considering
models not formulated in terms of field variables.
String theories have emerged as the most promising
candidates for unifying all fundamental interactions, gravity included;
it is therefore of high interest to investigate whether $CPT$-invariance
holds or not in such models. Indeed, these theories involve extended
objects as fundamental dynamical variables, which in the low energy
regime reduce to an infinite tower of field-like degrees of freedom.
Therefore, at least some of the requirements that makes $CPT$-invariance
unavoidable in quantum field theories do not hold anymore.

Despite these considerations, the general structure of the theory
could easily accommodate $CPT$-invariance,[16] and explicit investigations
also
support this possibility.[17] However, the knowledge of the vacuum structure
of string theory is not precise enough to draw definite conclusions.
Indeed, it has been proposed that in some string model a mechanism
for spontaneous breaking of Lorentz invariance can occur and, as a consequence,
lead to possible $CPT$-violating effects.[18] This phenomenon could originate
from certain type of interactions among tensor fields that are
known to be present in string field theory; if these interactions
destabilize the vacuum and produce non-trivial expectation values
for some of these tensor fields, than Lorentz symmetry is spontaneously
broken.

Even assuming that this mechanism is really at work, it is nevertheless
difficult to predict its effects in the low energy regime, at the level
of particle physics phenomena. At the present stage of string theory
knowledge, only an effective, phenomenological recipe can be given:
add by hand to the Lagrangian of the standard model $CPT$ and Lorentz
violating terms, with the additional requirement of maintaining as much
as possible its fundamental properties, like gauge invariance and
perturbative renormalizability.[19-21]
In this way, one obtains a phenomenological
extension of the standard theory that, although inspired by the fundamental
theory of strings, is effectively formulated in terms of ordinary field
theory. The model contains many unconstrained parameters, so that
in this formalism full predictivity is lost; nevertheless, at least
to a certain extent, one can still relate and classify $CPT$-violating effects
in different elementary particle systems.

For instance, the fermion sector of the standard model Lagrangian can
be augmented by terms of the form:
$$
{\cal L}^{CPT}_f=v^\mu\, \bar\psi\gamma_\mu\psi+
a^\mu\, \bar\psi\gamma_\mu\gamma_5\psi\ ;
\eqno(2.1)
$$
similarly, to the gauge sector one can add Chern-Simons like terms
(for simplicity, in the abelian case):
$$
{\cal L}^{CPT}_g=p^\mu\,\epsilon_{\mu\nu\lambda\rho} A^\nu F^{\lambda\rho}\ .
\eqno(2.2)
$$
In these expressions, $v^\mu$, $a^\mu$ and $p^\mu$ are (dimensionfull)
phenomenological constants; they are fixed vectors in space-time
({\it i.e.} constant external fields), so that invariance under (active)
Lorentz and $CPT$ transformations is broken by hand.

The new, additional pieces in (2.1) and (2.2) produce modifications of
various observables in many physical situations; the experiment can therefore
provide sensible bounds at least on some of the components of
$v^\mu$, $a^\mu$ and $p^\mu$. In particular, $v^\mu$ can be constrained
by experiments on neutral mesons systems,[22] while stringent bounds on
$p^\mu$ are given by the observation of radio waves coming from sources
at cosmological distance.[19]

The dynamics of extended objects (strings and branes) could give origin
to other $CPT$-violating phenomena, that can not be described in terms of
simple modifications of the basic standard model Lagrangian,
as in (2.1) and (2.2): these effects are thus
distinct from those discussed so far.
The basic idea is that, due to quantum fluctuations
and the appearance of virtual black holes, space-time should be
topological non-trivial at Planck's scale, and therefore it should be properly
described in terms of a ``foam'' structure;[23] as a consequence, this could
lead to possible loss of quantum coherence, which in turn could be accompanied
by $CP$ and $CPT$ violating effects.[24-29]

The proper framework to analyze these phenomena is that offered by
the study of open quantum systems.[30-32] Quite in general,
these systems can be modeled as being subsystems $S$ in interaction
with suitable, large environments $E$.
The total system $S+E$ is closed and therefore its dynamics can be
analyzed using standard quantum mechanical techniques:
it is realized by unitary operators, $U(t)=e^{-iH_{\rm tot} t}$,
generated by the total hamiltonian $H_{\rm tot}$, the sum of the
hamiltonian $H_S$ describing the dynamics of $S$ in absence of the
environment, the hamiltonian $H_E$ of $E$ and the interaction hamiltonian
$H_{SE}$. Nevertheless, the time evolution of the subsystem $S$ alone,
obtained by eliminating the environment degrees of freedom,
usually develops dissipation and irreversibility.

In the language of density matrices, the reduced dynamics is obtained by
tracing over the environment variables:
$$
\rho_S(0)\mapsto \rho_S(t)={\rm Tr}_E\Big[U(t)\, \rho_{S+E}\, U^\dagger(t)
\Big]\ ,\eqno(2.3)
$$
where the density matrix $\rho_S$ describes the state of $S$, while
$\rho_{S+E}$ represents the initial state of the total system;
in absence of initial correlations between $S$ and $E$, a situation commonly
encountered in many physical applications, it can be written in factorized
form: $\rho_{E+S}=\rho_S\otimes\rho_E$.

In general, the dynamics given in (2.3) is very complicated and can not be
explicitly described; however, when the interaction between
system and environment is ``weak'',
a condition usually very well satisfied in practice,
it becomes free from memory effects and can be realized in terms
of linear maps. In addition, these transformations are seen
to satisfy very basic physical requirements, like forward in time
composition (semigroup property), probability conservation, entropy increase,
complete positivity. They form a so-called quantum dynamical semigroup.

This description is very general and has been applied to successfully
model a large variety of physical situations; in particular,
it has been used to study quantum statistical systems,[30-32] to analyze
dissipative effects in quantum optics,[33-35] to describe the interaction
of a microsystem with a macroscopic measuring apparatus.[36, 37] As mentioned
before, it is also the natural framework to study phenomena leading
to violation of $CPT$-symmetry induced by dissipative effects.[38]

More in general, effective time evolutions based on quantum dynamical
semigroups
can also accommodate more ``standard'' $CPT$-breaking terms, of the kind
induced by the terms in (2.1) and (2.2):[39, 40]
a unified, phenomenological description
of $CPT$-violating phenomena is therefore at hand.

%\vskip 2cm
\vfill\eject

\noindent
\line{{\bf 3. PHENOMENOLOGY OF {\script CPT }-VIOLATING EFFECTS}\hfill}
\line{\phantom{\bf 3. }{\bf IN THE NEUTRAL KAON SYSTEM}\hfill}

\bigskip
\noindent
One of the most promising venues to look for $CPT$-violating effects
is certainly the system of neutral kaons: the sofistication of present
(and planned) kaon experiments is so high that systematic and very accurate
searches are possible. Although a predictive, microscopic theory
for such effects is presently unavailable, a phenomenological
approach based on quantum dynamical semigroups can nevertheless provide
a sufficiently general framework for a meaningful interpretation of such
non-standard phenomena.

The evolution and decay of the neutral kaon system can be effectively modeled
by means of a two-dimensional Hilbert space.[41] The states of definite $CP$,
$$
|K_1\rangle={1\over\sqrt{2}}\Big[|K^0\rangle+|\overline{K^0}\rangle\Big]
\ ,\quad
|K_2\rangle={1\over\sqrt{2}}\Big[|K^0\rangle-|\overline{K^0}\rangle\Big]\ ,
\eqno(3.1)
$$
constitute an orthonormal basis in this space.
In this basis, a kaon state can be conveniently described
by means of a density matrix $\rho$,
$$
\rho=\left(\matrix{
\rho_1&\rho_3\cr
\rho_4&\rho_2}\right)\ , \qquad \rho_4\equiv\rho_3^*\ ;
\eqno(3.2)
$$
this is a positive hermitian operator, {\it i.e.} with real, positive
eigenvalues, and constant trace (in case of unitary evolutions).

The evolution in time of the kaon system can then be formulated in
terms of a linear equation for $\rho$; it takes the general form:[38, 39]
$$
{\partial\rho(t)\over \partial t}= -i H_{\rm eff}\ \rho(t)+i\rho(t)\,
H_{\rm eff}^\dagger + L[\rho(t)]\ .\eqno(3.3)
$$
The first two terms on the r.h.s. of this equation are the standard quantum
mechanical ones: they contain the effective hamiltonian
$H_{\rm eff}=M-i{\mit\Gamma}/2$, which includes a non-hermitian part,
characterizing the natural width of the kaon states.

The entries of this matrix can be expressed in terms of the complex parameters
$\epsilon_S$, $\epsilon_L$, appearing in the eigenstates of
$H_{\rm eff}$,
$$
|K_S\rangle={1\over(1+|\epsilon_S|^2)^{1/2}}
\left(\matrix{1\cr\epsilon_S}\right)\ ,\qquad
|K_L\rangle={1\over(1+|\epsilon_L|^2)^{1/2}}
\left(\matrix{\epsilon_L\cr 1}\right)\ ,
\eqno(3.4)
$$
and the four real parameters
$m_S$, $\gamma_S$ and $m_L$,
$\gamma_L$, the masses and widths of the states in (3.4),
characterizing the eigenvalues of $H_{\rm eff}$:
$\lambda_S=m_S-{i\over 2}\gamma_S$, $\lambda_L=m_L-{i\over 2}\gamma_L$.[42]
It proves convenient to use also the following positive combinations:
$\Delta\Gamma=\gamma_S-\gamma_L$, $\Delta m=m_L-m_S$,
as well as the complex quantities:
$\Gamma_\pm=\Gamma\pm i \Delta m$ and
$\Delta\Gamma_\pm=\Delta\Gamma\pm 2i\Delta m$,
with $\Gamma=(\gamma_S+\gamma_L)/2$.
One easily checks that $CPT$-invariance is broken when
$\epsilon_S\neq\epsilon_L$, while a nonvanishing
$\epsilon_S=\epsilon_L$ implies violation of $CP$ symmetry.
For these reasons, in order to parametrize these so-called ``indirect''
violations, the constants $\epsilon_K=(\epsilon_S+\epsilon_L)/2$
and $\delta_K=(\epsilon_S-\epsilon_L)/2$ are sometimes introduced.

The various paradigms for $CPT$-violation mentioned in the previous section
can provide, at least in principle, some informations on these parameters.
For instance, the effective approach to $CPT$-violation motivated by a possible
``stringy'' spontaneous symmetry breaking mechanism can have some bearings
on the parameter $\delta_K$, via contributions induced by
the first term in (2.1); it predicts a dependence of $\delta_K$
on differences of the parameters $v^\mu$
pertaining to different quark flavours and on the kaon 4-velocity.[22]

On the other hand, the additional piece $L[\rho]$ in the evolution equation
(3.3) takes care of a different kind of phenomena leading to possible
violations of $CP$ and $CPT$ symmetries; it encodes effects leading to
dissipation and irreversibility that can not be reproduced by the
standard quantum mechanical treatments, like those just mentioned.

The explicit form of the linear map $L[\rho]$ can be uniquely fixed
by taking into account the basic physical requirements that the complete
time evolution, $\lambda_t: \rho(0)\mapsto\rho(t)$, generated
by (3.3) needs to satisfy; the one-parameter family of maps
$\lambda_t$ must transform density matrices into density matrices
and have the properties of increasing the von Neumann entropy,
$S=-{\rm Tr}[\rho\ln\rho]$, of obeying the semigroup composition law,
$\lambda_t[\rho(t')]=\rho(t+t')$, for $t,\, t'\geq0$, of being completely
positive. If one expands the $2\times 2$ matrix $\rho$ in terms of
Pauli matrices, $\sigma_1$, $\sigma_2$, $\sigma_3$, and the identity,
$\sigma_0$, $\rho=\sigma_\mu\rho_\mu$, $\mu=0,1,2,3$,
$L[\rho]$ can be represented by a symmetric $4\times 4$ matrix
$\big[L_{\mu\nu}\big]$,
acting on the column vector with components $(\rho_0,\rho_1,\rho_2,\rho_3)$.
It can be parametrized by the six real
constants $a$, $b$, $c$, $\alpha$, $\beta$, and $\gamma$:[39]
$$
\big[L_{\mu\nu}\big]=-2\left(\matrix{0&0&0&0\cr
                                     0&a&b&c\cr
                                     0&b&\alpha&\beta\cr
                                     0&c&\beta&\gamma\cr}\right)
\ ,\eqno(3.5)
$$
with $a$, $\alpha$ and $\gamma$ non-negative.
These parameters are not all independent; to assure the
complete positivity of the time-evolution
generated by (3.3), they have to satisfy the following inequalities:
$$
\eqalign{
&2\,R\equiv\alpha+\gamma-a\geq0\ ,\cr
&2\,S\equiv a+\gamma-\alpha\geq0\ ,\cr
&2\,T\equiv a+\alpha-\gamma\geq0\ ,\cr
&RST-2\, bc\beta-R\beta^2-S c^2-T b^2\geq 0\ .
}\hskip -1cm
\eqalign{
&RS-b^2\geq 0\ ,\cr
&RT-c^2\geq 0\ ,\cr
&ST-\beta^2\geq 0\ ,\cr
&\phantom{\beta^2}\cr
}\eqno(3.6)
$$

As already mentioned, although the basic general idea behind such an approach
to the kaon dynamics is that quantum phenomena at Planck's scale
produce loss of phase-coherence, it should be stressed that the form
(3.3), (3.5) of the kaon time-evolution is quite independent from the
microscopic mechanism responsible for the dissipative effects;
indeed, in view of the properties mentioned above, the evolution of any
quantum irreversible process can be effectively modeled using
quantum dynamical semigroups. In this respect, the equations (3.3), (3.5)
should be regarded as phenomenological in nature, and therefore
quite suitable to experimental test: any signal of non-vanishing
value for some of the parameters in (3.5) would certainly attest
in a model independent way the presence of non-standard, dissipative
effects in kaon physics.

Among the physical requirements that the evolution (3.3), (3.5)
satisfies, complete positivity
is perhaps the less intuitive. Indeed, it has not been
enforced in many analysis, in favor of the more obvious
simple positivity.
Simple positivity is in fact generally enough to guarantee that
the eigenvalues of the kaon density matrix $\rho(t)$
remain positive at any time; this requirement is obviously crucial
for the consistency of the formalism, in view of the interpretation of the
eigenvalues of $\rho(t)$ as probabilities.

Complete positivity is a stronger property, since
it assures the positivity of the density matrix describing the
states of a larger system, obtained by coupling in a trivial
way the neutral kaon system with another arbitrary
finite-dimensional one. Although trivially satisfied by standard
quantum mechanical (unitary) time-evolutions,
the requirement of complete positivity
seems at first a mere technical complication.
Nevertheless, it turns out to be essential in properly
treating correlated systems, like the two neutral kaons
coming from the decay of $\phi$-meson (a common physical situation
at $\phi$-factories); it assures the absence of
unphysical effects, like the appearance of negative probabilities,
that could occur for just simply positive dynamics.[43]

An example is provided by the extended kaon dynamics originally discussed
in [24], and further developed in [44, 45];
there, the generator of the dissipative part of the evolution is
taken of the form (3.5), but with $a=b=c=\,0$, $\alpha\neq\gamma$,
and $\alpha\gamma\geq\beta^2$, to assure simple positivity.
One can check that the total time evolution is no longer
completely positive: the inequalities (3.6) are clearly violated;
and indeed, in the case of correlated kaons, one can easily evidentiate
physical inconsistencies for such simplified dynamics: some
expectation values of the corresponding two-kaon density matrix
become negative.[43] The only way to avoid this serious drawback is to restore
complete positivity, by letting $\alpha=\gamma$ and $\beta=\,0$,
a consequence of (3.6) for a vanishing $a$.%
\footnote{$^\dagger$}{As a result of recent investigations on the effective
D-brane dynamics, a non-linear time evolution is suggested as a possible
alternate way of avoiding these pathologies;[46] in this respect, we note
that the correct treatment of non-linear dynamics for correlated systems is
still an unsolved problem.[47]}
Let us stress that these considerations
are not purely academic: the need of complete positivity in the neutral kaon
dynamics can be actually probed using experimental set-ups at
$\phi$-factories (for a complete discussion, see [48]).

Physical observables of the neutral kaon system can be computed from the
density matrix $\rho(t)$; its time evolution is given as a solution of the
equation (3.3), (3.5) for a given initial state $\rho$.
On the basis of rough dimensional estimates,[44, 45, 38] the parameters
$a$, $b$, $c$, $\alpha$, $\beta$ and $\gamma$ appearing in (3.5) are
expected to be very small (at most of order $m_K^2/M_P\sim 10^{-19}\ {\rm
GeV}$,
with $m_K$, $M_P$ the kaon and Planck's mass), roughly of
the same order of magnitude of
$\epsilon_S\Delta\Gamma$ and $\epsilon_L\Delta\Gamma$; therefore, one can use
an expansion in all these small parameters, so that approximate
expressions for the entries of $\rho(t)$ can be explicitly
obtained.[39, 40] In particular, one can work
out the contributions $\rho_L$ and $\rho_S$
that correspond to the physical ``long lived'' and ``short lived''
neutral kaon states:
$$
\rho_L=\left[\matrix{
\left|\epsilon_L+{2C^*\over\Delta\Gamma_-}\right|^2+{\gamma\over\Delta\Gamma}
-8\left|{C\over\Delta\Gamma_+}\right|^2
-4\,{\cal R}e\left({\epsilon_L C\over\Delta\Gamma}\right)
& \epsilon_L+{2 C^*\over\Delta\Gamma_-}\cr
\phantom{.}&\phantom{.}\cr
\epsilon_L^*+{2 C\over\Delta\Gamma_+} & 1}\right]\ ,
\eqno(3.7a)
$$
and
$$
\rho_S=\left[\matrix{ 1 &
\epsilon_S^*+{2 C^*\over\Delta\Gamma_+}\cr
\phantom{.}&\phantom{.}\cr
\epsilon_S+{2 C\over\Delta\Gamma_-} &
\left|\epsilon_S+{2C\over\Delta\Gamma_-}\right|^2 -{\gamma\over\Delta\Gamma}
-8\left|{C\over\Delta\Gamma_+}\right|^2
-4\,{\cal R}e\big({\epsilon_S C^*\over\Delta\Gamma}\big)}\right]\ ,
\eqno(3.7b)
$$
where
$$
C=c+i\beta\ .\eqno(3.8)
$$
Note that the states $\rho_L$ and $\rho_S$ are mixed: the matrices
(3.7) are not projectors; only in absence of the contribution
in (3.5) one recovers the matrices $|K_L\rangle\langle K_L|$ and
$|K_S\rangle\langle K_S|$.

As a further remark, let us point out that in writing down the evolution
equation (3.3), an arbitrary phase convention for the states
$|K^0\rangle$, $|\overline{K^0}\rangle$, and $|K_S\rangle$, $|K_L\rangle$
has been implicitly fixed. We remark that physical observables, being the
result
of a trace operation (see below), are by definition independent from any
phase convention. Provided a complete set of observables are computed,
one can always consistently re-express a result obtained in a given phese
convention, into any other phase choice. Nevertheless, once a phase
choice is adopted and a given approximation used, one should consistently
stick within this convention, paying attention not to generalize conclusions
drawn from only a limited number of physical observations.
In particular, as already observed, the constants
$a$, $b$, $c$, $\alpha$, $\beta$ and $\gamma$ appearing in the dissipative
part (3.5) of the dynamical equation (3.3) are expected to by tiny.
Since the dependence of the map $L[\rho]$ on these parameters is linear,
the latter, being small
in one phase convention, remain small in any other phase choice.
As a consequence, the generality of the discussion in the next sections
is by no mean affected by the selection of any particular phase convention.

\vskip 2cm

\noindent
{\bf 4. SINGLE NEUTRAL KAON OBSERVABLES}

\bigskip
\noindent
The evolution equation (3.3), (3.5) provides the most
general linear dynamics for the neutral kaon system,
encoding $CP$ and $CPT$-violating effects.
These are the so-called ``indirect'' effects: further
``direct'' violations can occur. They are related to
specific observables, typically those associated with the
decays of the kaons into pion states, and into
semileptonic states.
In the language of density matrices, these final decay states
are described by suitable hermitian operators; taking the trace
of these operators with $\rho(t)$ allows computing the explicit
time dependence of various experimentally
relevant observables.[44, 45, 39, 40, 49]

For instance, the operators ${\cal O}_{+-}$ and
${\cal O}_{00}$ that describe the
$\pi^+\pi^-$ and $2\pi^0$ final states have the form:
$$
{\cal O}_{+-}\sim\left[\matrix{1&r_{+-}\cr
                              r_{+-}^*&|r_{+-}|^2\cr}\right]\ ,\qquad
{\cal O}_{00}\sim\left[\matrix{1&r_{00}\cr
                              r_{00}^*&|r_{00}|^2\cr}\right]\ .
\eqno(4.1)
$$
To lowest order in all small parameters, the complex
constants $r_{+-}$ and $r_{00}$, can be written as:
$$
r_{+-}=\varepsilon-\epsilon_L+\varepsilon^\prime\ ,\qquad
r_{00}=\varepsilon-\epsilon_L-2\varepsilon^\prime\ ,
\eqno(4.2)
$$
where the parameters $\varepsilon$ and $\varepsilon'$ take the familiar
expressions:[42]
$$
\varepsilon=\bigg[{\epsilon_L+\epsilon_S\over2} +
i\, {{\cal I}m(A_0)\over {\cal R}e(A_0)}\bigg]+
\bigg[{\epsilon_L-\epsilon_S\over2} +
{{\cal R}e(B_0)\over {\cal R}e(A_0)}\bigg]\ ,\eqno(4.3)
$$
and
$$
\varepsilon^\prime= {i e^{i(\delta_2-\delta_0)}\over\sqrt2}\,
{{\cal R}e(A_2)\over {\cal R}e(A_0)}\, \bigg[
{{\cal I}m(A_2)\over {\cal R}e(A_2) }
-{{\cal I}m(A_0)\over {\cal R}e(A_0) }\bigg]
+i\bigg[{{\cal R}e(B_0)\over  {\cal R}e(A_0) }
-{ {\cal R}e(B_2)\over {\cal R}e(A_0)}\bigg]\ .\eqno(4.4)
$$
In these formulas, the $s$-wave phase shifts $\delta_i$ and
the complex amplitudes $A_i$, $B_i$, $i=0,2$, have been introduced;
they appear in the usual parametrization of the amplitudes for the decay
of a neutral kaon into two pions:
$$
\eqalignno{
&{\cal A}(K^0\rightarrow \pi^+\pi^-)=(A_0+B_0)\, e^{i\delta_0}+{1\over\sqrt2}\,
(A_2+B_2)\, e^{i\delta_2}\ ,&(4.5a)\cr
&{\cal A}(K^0\rightarrow 2\pi^0)=(A_0+B_0)\, e^{i\delta_0}-\sqrt{2}\,
(A_2+B_2)\, e^{i\delta_2}\ ;&(4.5b)}
$$
the indices 0, 2 refers to the total isospin. (The amplitudes for
the $\overline{K^0}$ decays are obtained from (4.5) with the substitutions:
$A_i\rightarrow A_i^*$ and $B_i\rightarrow -B_i^*$.)
The imaginary parts of $A_i$ signals direct $CP$-violation, while a non zero
value for $B_i$ will also break $CPT$-invariance.
As a consequence, in the expressions (4.3) and (4.4),
the first, second square brackets contain the $CP$,
respectively $CPT$, violating parameters arising both
from the effective hamiltonian $H_{\rm eff}$,
as well as from their decay amplitudes.%
\footnote{$^\dagger$}{The factor
$\omega={\cal R}e(A_2)/{\cal R}e(A_0)$ corresponds to the suppression due to
the $\Delta I=1/2$ rule in the kaon two-pion decays. This ratio is known to be
small; therefore, in presenting the above formulas, first order terms in the
small parameters multiplied by $\omega^2$ have been consistently neglected.}

Results similar to those presented in (4.1) are obtained for the
matrices ${\cal O}_{\pi^+\pi^-\pi^0}$,
${\cal O}_{3\pi^0}$, ${\cal O}_{\ell^-}$ and ${\cal O}_{\ell^+}$
that describe the decays into $\pi^+\pi^-\pi^0$, $3\pi^0$,
$\pi^+\ell^-\bar\nu$ and $\pi^-\ell^+\nu$; for explicit expressions, see
Refs.[40, 49].

With the help of these matrices, one can compute
the time-dependence
of various useful observables of the neutral kaon system.
For example, in the case of charged pions,
the decay rate for a pure $K^0$ initial state $\rho_{K^0}$ is:
$$
\eqalign{
R_{+-}(t)&={
{\rm Tr}\Big[\rho_{K^0}(t){\cal O}_{+-}\Big]\over
{\rm Tr}\Big[\rho_{K^0}(0){\cal O}_{+-}\Big]}\cr
&=e^{-\gamma_S t}+R_{+-}^L\, e^{-\gamma_L t}+
2\, e^{-\Gamma t}\, |\eta_{+-}|\cos(\Delta m\, t-\phi_{+-})\ ,}
\eqno(4.6)
$$
where $R_{+-}^L$ is the two-pion decay rate for the $K_L$ state:
$$
R_{+-}^L=
\left|\epsilon_L+{2 C^*\over\Delta\Gamma_-}+r_{+-}\right|^2
+{\gamma\over\Delta\Gamma}-8\left|{C\over\Delta\Gamma_+}\right|^2
-4\, {\cal R}e\left({\epsilon_L C\over\Delta\Gamma}\right)\ ,
\eqno(4.7)
$$
while the interference term is determined by the combination
$$
\epsilon_L-{2C^*\over\Delta\Gamma_-}+r_{+-}
\equiv\eta_{+-}=|\eta_{+-}|\, e^{i\phi_{+-}}\ .
\eqno(4.8)
$$

Other interesting observables, directly measured in experiments,
are the asymmetries associated with the decay into the final state $f$
of an initial $K^0$, described by the density matrix
$\rho_{K^0}$, as compared to the corresponding decay
into the conjugate state $\bar f$ of an initial
$\overline{K^0}$, described by the matrix $\rho_{\bar K^0}$.
All these asymmetries have the general form
$$
A(t)={
{\rm Tr}\Big[\rho_{\bar{K^0}}(t){\cal O}_{\bar f}\Big]-
\Big[\rho_{K^0}(t){\cal O}_f\Big]\over
{\rm Tr}\Big[\rho_{\bar{K^0}}(t){\cal O}_{\bar f}\Big]+
\Big[\rho_{K^0}(t){\cal O}_f\Big]}\ .
\eqno(4.9)
$$

The phenomenological quantities $R^L_{+-}$ and $\eta_{+-}$
are accessible to the experiment,
so that they can be used, together with other asymmetries
involving three-pion and semileptonic final states,
to obtain estimates on $CPT$-violating phenomena.
Of particular interest are the bounds that can be derived
on the dissipative effects, encoded in (3.5);
using the most recent results,[50, 5] one can indeed obtain
some information on some of the non-standard parameters:
$$
\eqalign{
&a=(2.5 \pm 2.6)\times 10^{-17}\ \hbox{GeV}\ ,\cr
&c=(0.7 \pm 1.2)\times 10^{-17}\ \hbox{GeV}\ ,\cr
&\alpha=(1.8\pm 4.4)\times 10^{-17}\ \hbox{GeV}\ ,\cr
&\beta=(-0.7 \pm 1.3)\times 10^{-17}\ \hbox{GeV}\ ,\cr
&\gamma=(0.1 \pm 22.0)\times 10^{-20}\ \hbox{GeV}\ .\cr}
\eqno(4.15)
$$
Unfortunately, the precision of the present experimental
results on single neutral kaons is not high enough to allow a
meaningful test of the hypothesis of complete positivity.
Although more complete and precise data will surely be available
in the future, the most promising venues for studying
the consequences of the dissipative dynamics in (3.3), (3.5)
are certainly the experiments
on correlated neutral kaons at $\phi$-factories.

\vskip 2cm

\noindent
{\bf 5. CORRELATED NEUTRAL KAON OBSERVABLES}

\bigskip
\noindent
The time evolution of a system of two correlated neutral kaons
coming from the decay of a $\phi$-meson can be discussed
using the results on the dynamics of a single kaon presented
in the previous section.[40]

Since the $\phi$-meson has spin one, its decay into two spinless bosons
produces
an antisymmetric spatial state. In the $\phi$ rest frame, the two neutral kaons
are produced flying apart with opposite momenta; in the basis $|K_1\rangle$,
$|K_2\rangle$, the resulting state can thus be described by:
$$
|\psi_A\rangle= {1\over\sqrt2}\Big(|K_1,-p\rangle \otimes  |K_2,p\rangle -
|K_2,-p\rangle \otimes  |K_1,p\rangle\Big)\ .
\eqno(5.1)
$$
The corresponding density operator $\rho_A$ is a $4\times 4$ matrix
and its time-evolution can be obtained by assuming that, once
produced in a $\phi$ decay, the kaons evolve in time each according to the
completely positive map generated by the equation (3.3).%
\footnote{$^\dagger$}{
This assures that the resulting evolution map
$\rho_A\mapsto\rho_A(t)$ is
completely positive and of semigroup type; further, this dynamics
is independent from the particular situation under study and can be easily
generalized to systems containing more than two particles.
Although other possibilities are conceivable, this choice is the most
natural one. In fact, it is very hard to produce linear dynamical maps
for a system of two particles not in factorized form
without violating very basic physical principles. Indeed, if one requires that
after tracing over the degrees of freedom of one particle, the resulting
time evolution for the remaining one be completely positive,
of semigroup type and independent from
the initial state of the first particle, than the only
natural possibility is a factorized dynamics.}

The typical observables that can be studied in such physical situations
are double decay rates, {\it i.e.} the probabilities that a kaon decays
into a final state $f_1$ at proper time $\tau_1$, while the other kaon
decays into the final state $f_2$ at proper time $\tau_2$:
$$
{\cal G}(f_1,\tau_1; f_2,\tau_2)\equiv
\hbox{Tr}\Big[\Big({\cal O}_{f_1}\otimes{\cal O}_{f_2}\Big)
\rho_A(\tau_1,\tau_2)\Big]\ ;
\eqno(5.2)
$$
as before, the operators ${\cal O}_{f_1}$ and ${\cal O}_{f_2}$
are the $2\times 2$ hermitian matrices describing the final decay states.

As an example, let us consider the situation in which $\tau_1=\tau_2=\tau$
and the two kaons decay into the same final state $\pi^+\pi^-$:
$$
\eqalign{
{\cal G}(&\pi^+ \pi^-,\tau;\pi^+\pi^-,\tau)\sim
e^{-\gamma_S\tau}
\bigg\{ e^{-\gamma_L\tau}\Big(R^L_{+-}-|\eta_{+-}|^2\Big)\cr
&-e^{-\gamma_S\tau}\bigg[{\gamma\over\Delta\Gamma}
+8\left|{C\over\Delta\Gamma_+}\right|^2-4\,{\cal R}e\bigg({\epsilon_L\,
C\over\Delta\Gamma}\bigg)\bigg]
-e^{-\Gamma\tau}\, 8\, {\cal R}e\bigg({\eta_{+-}\,
C\over\Delta\Gamma_+}
e^{-i\Delta m\tau}\bigg)\bigg\}\ .}
\eqno(5.3)
$$
This time-behaviour is completely different from the one required by
ordinary quantum mechanics, which predicts:
$$
{\cal G}(f,\tau;f,\tau)\equiv 0\ ,
\eqno(5.4)
$$
for all final states $f$, due to the antisymmetry of the initial state
$\rho_A$.
Therefore, by studying double decay rates in a high-luminosity
$\phi$-factory it will be possible to determine the values of the non-standard
parameters $a$, $b$, $c$, $\alpha$, $\beta$, $\gamma$.
For instance, the long time behaviour
$(\tau\gg 1/\gamma_S)$ of the three pion probability
gives direct informations on the parameter $\gamma$:
$$
{\cal G}(\pi^+\pi^-\pi^0,\tau\,;\pi^+\pi^-\pi^0,\tau)\sim
{\gamma\over\Delta\Gamma}\ e^{-2\gamma_L\tau}\ .
\eqno(5.5)
$$
Similarly, the small time behaviour of the ratio of semileptonic probabilities,
$$
{{\cal G}(\ell^\pm,\tau;\ell^\pm,\tau)\over
{\cal G}(\ell^\pm,\tau;\ell^\mp,\tau)}\sim
2\ a\ \tau\ ,
\eqno(5.6)
$$
is sensible to the parameter $a$.

However, much of the analysis at $\phi$-factories is carried out using
integrated distributions at fixed time interval $\tau=\tau_1-\tau_2$.[51]
One then deals with single-time distributions, defined by:
$$
{\mit\Gamma}(f_1,f_2;\tau)=\int_0^\infty dt\, {\cal G}(f_1,t+\tau;f_2,t)\ ,
\qquad \tau>0\ .\eqno(5.7)
$$
Starting with these integrated probabilities, one can form asymmetries
that are sensitive to various parameters in the theory.
A particularly interesting example is given by the following observable,
involving two-pion final states,
$$
{\cal A}_{\varepsilon^\prime}(\tau)={
{\mit\Gamma}(\pi^+\pi^-,2\pi^0;\tau) -
{\mit\Gamma}(2\pi^0,\pi^+\pi^-;\tau)\over
{\mit\Gamma}(\pi^+\pi^-,2\pi^0;\tau) +
{\mit\Gamma}(2\pi^0,\pi^+\pi^-;\tau) }\ ;
\eqno(5.8)
$$
it is used for the determination of the ratio
$\varepsilon'/\varepsilon$.
Indeed, one can show that, to first order in all small parameters:
$$
{\cal A}_{\varepsilon^\prime}(\tau)=3\,
{\cal R}e\Big({\varepsilon^\prime\over\varepsilon}\Big)\
{N_1(\tau)\over D(\tau)}
-3\, {\cal I}m\Big({\varepsilon^\prime\over\varepsilon}\Big)\
{N_2(\tau)\over D(\tau)}\ ,\eqno(5.9)
$$
where the $\tau$-dependent coefficients $N_1$, $N_2$ and $D$ are now
functions of $\varepsilon$ and of the
dissipative parameters $c$, $\beta$ and $\gamma$ (for
explicit expressions, see [52].)

The clear advantage of using the asymmetry
${\cal A}_{\varepsilon^\prime}$ to determine
the value of $\varepsilon'/\varepsilon$ in comparison to the
familiar ``double ratio'' method,[53, 54] is that, at least in principle,
both real and imaginary
part can be extracted from the time behaviour of (5.9).
Due to the presence of the dissipative parameters however,
this appears to be much more problematic than in the
standard case; a meaningful determination of
$\varepsilon'/\varepsilon$ is possible provided independent estimates
on $c$, $\beta$ and $\gamma$ are obtained from
the measure of other independent asymmetries.
This is particularly evident if one looks at the large-time
limit $(\tau\gg1/\gamma_S)$ of (5.9):
$$
\eqalign{
{\cal A}_{\varepsilon^\prime}(\tau)
\sim
3\,{\cal R}e\left({\varepsilon'\over\varepsilon}\right)\,&
{ |\varepsilon|^2+
2\,{\cal R}e\big(\varepsilon\, C/\Delta\Gamma_+\big)\over |\varepsilon|^2
+{\cal D} }\cr
&\hskip 2 cm
-6\,{\cal I}m\left({\varepsilon'\over\varepsilon}\right)\,
{ {\cal I}m\big(\varepsilon\, C/\Delta\Gamma_+\big)\over|\varepsilon|^2
+{\cal D} }\ ,\cr}\eqno(5.10)
$$
where
$$
{\cal D}={\gamma\over\Delta\Gamma}-4\left|{C\over\Delta\Gamma_+}\right|^2
+4\, {\cal R}e\left({\varepsilon\, C\over\Delta\Gamma_+}\right)
-4\, {\cal R}e\left({\epsilon_L C\over\Delta\Gamma}\right)\ ;\eqno(5.11)
$$
only when $c=\beta=\gamma=\,0$, the expression in (5.10) reduces to the
standard result:
${\cal A}_{\varepsilon^\prime}\sim 3\,{\cal R}e(\varepsilon'/\varepsilon)$.
Therefore, if the non-standard, dissipative parameters in (3.5)
are found to be non-zero, even neglecting the contribution from
the imaginary part, the actual value of
${\cal R}e(\varepsilon'/\varepsilon)$
could be significantly different from the measured value of the
quantity ${\cal A}_{\varepsilon^\prime}/3$.[52]

In conclusion, dissipative effects in the dynamics of both single
and correlated neutral kaon systems could affect the precise determination
of various relevant quantities in kaon physics;
dedicated experiments, in particular those involving correlated kaons,
will certainly provide stringent bounds on these
dissipative effects, thus clarifying their role in generating
$CP$ and $CPT$-violating phenomena.

\vskip 3cm
%\vfill\eject

\centerline{\bf REFERENCES}
\bigskip\medskip

\item{1.} R. Jost, {\it The General Theory of Quantized Fields},
(American Mathematical Society, Providence, 1965)
\smallskip
\item{2.} R.F. Streater and A.S. Wightman {\it PCT, Spin and Statistics,
and All That}, (Benjamin-Cummings, New York, 1968)
\smallskip
\item{3.} F. Strocchi, {\it Selected Topics on the General Properties
of Quantum Field Theory}, (World Scientific, Singapore, 1993)
\smallskip
\item{4.} A.S. Wightman, The general theory of quantized fields in
the 1950's, in {\it Pions and Quarks}, L.M. Brown, M. Dresden and
L. Hoddesdon, eds., (Cambridge University Press, Cambridge, 1989)
\smallskip
\item{5.} Particle Data Group, Eur. Phys. J. C {\bf 3} (1998) 1
\smallskip
\item{6.} N.P. Lipshutz, Phys. Rev. {\bf 144} (1966) 1300
\smallskip
\item{7.} R. Carosi {\it et al.}, Phys. Lett. {\bf B237} (1990) 303
\smallskip
\item{8.} L. Lavoura and J.P. Silva, Phys. Rev. D {\bf 60} (1999) 056003
\smallskip
\item{9.} A.I. Sanda, Comments on $CPT$ tests, {\tt hep-ph/9902353}
\smallskip
\item{10.} N.N. Bogolubov, A.A. Logunov, A.I. Oksak and I.T. Todorov,
{\it General Principles of Quantum Field Theory},
(Kluwer Academic Publishers, Dordrecht, 1990), Appendix I
\smallskip
\item{11.} E. Majorana, Nuovo Cim. {\bf 9} (1932) 335
\smallskip
\item{12.} O.W. Greenberg, Phys. Lett. {\bf B416} (1998) 144
\smallskip
\item{13.} A.I. Oksak and I.T. Todorov, Comm. Math. Phys. {\bf 11} (1968) 125
\smallskip
\item{14.} N.N. Bogolubov, A.A. Logunov and I.T. Todorov,
{\it Introduction to Axiomatic Quantum Field Theory},
(Benjamin, Reading, 1975), Ch. 9
\smallskip
\item{15.} P. Carruthers, J. Math. Phys. {\bf 9} (1968) 928, 1835;
Phys. Rev. {\bf 172} (1968) 1406
\smallskip
\item{16.} J. Polchinski, {\it String Theory},
({Cambridge University Press, Cambridge, 1998})
\smallskip
\item{17.} A. Pasquinucci and K. Roland, Nucl. Phys. {\bf B473} (1996) 31;
{\it ibid.} {\bf B485} (1997) 241; Erratum, {\it ibid.} {\bf 494} (1997) 486
\smallskip
\item{18.} V.A. Kostelecky and S. Samuel, Phys. Rev. D {\bf 39} (1989) 683
\smallskip
\item{19.} S. Carroll, G. Field and R. Jackiw, Phys. Rev. D {\bf 41}
(1990) 1231
\smallskip
\item{20.} S. Coleman and S.L. Glashow, Phys. Lett. {\bf B405} (1997) 249
\smallskip
\item{21.} D. Colladay and V.A. Kostelecky, Phys. Rev. D {\bf 58} (1998) 116002
\smallskip
\item{22.} V.A. Kostelecky, Phys. Rev. Lett. {\bf 80} (1998) 1818
\smallskip
\item{23.} S. Hawking, Comm. Math. Phys. {\bf 87} (1983) 395; Phys. Rev. D
{\bf 37} (1988) 904; Phys. Rev. D {\bf 53} (1996) 3099;
S. Hawking and C. Hunter, Phys. Rev. D {\bf 59} (1999) 044025
\smallskip
\item{24.} J. Ellis, J.S. Hagelin, D.V. Nanopoulos and M. Srednicki,
Nucl. Phys. {\bf B241} (1984) 381;
\smallskip
\item{25.} S. Coleman, Nucl. Phys. {\bf B307} (1988) 867
\smallskip
\item{26.} S.B. Giddings and A. Strominger, Nucl. Phys. {\bf B307} (1988) 854
\smallskip
\item{27.} M. Srednicki, Nucl. Phys. {\bf B410} (1993) 143
\smallskip
\item{28.} W.G. Unruh and R.M. Wald, Phys. Rev. D {\bf 52} (1995) 2176
\smallskip
\item{29.} L.J. Garay, Phys. Rev. Lett. {\bf 80} (1998) 2508;
Phys. Rev. D {\bf 58} (1998) 124015
\smallskip
\item{30.} R. Alicki and K. Lendi, {\it Quantum Dynamical Semigroups and
Applications}, Lect. Notes Phys. {\bf 286}, (Springer-Verlag, Berlin, 1987)
\smallskip
\item{31.} V. Gorini, A. Frigerio, M. Verri, A. Kossakowski and
E.C.G. Surdarshan, Rep. Math. Phys. {\bf 13} (1978) 149
\smallskip
\item{32.} H. Spohn, Rev. Mod. Phys. {\bf 53} (1980) 569
\smallskip
\item{33.} W.H. Louisell, {\it Quantum Statistical Properties of Radiation},
(Wiley, New York, 1973)
\smallskip
\item{34.} C.W. Gardiner, {\it Quantum Noise}, (Springer, Berlin, 1992)
\smallskip
\item{35.} M.O. Scully and M.S. Zubairy,
{\it Quantum Optics}, (Cambridge University Press, Cambridge, 1997)
\smallskip
\item{36.} L. Fonda, G.C. Ghirardi and A. Rimini, Rep. Prog. Phys.
{\bf 41} (1978) 587
\smallskip
\item{37.} H. Nakazato, M. Namiki and S. Pascazio,
Int. J. Mod. Phys. {\bf B10} (1996) 247
\smallskip
\item{38.} F. Benatti and R. Floreanini, Ann. of Phys. {\bf 273} (1999) 58
\smallskip
\item{39.} F. Benatti and R. Floreanini, Nucl. Phys. {\bf B488} (1997) 335
\smallskip
\item{40.} F. Benatti and R. Floreanini, Nucl. Phys. {\bf B511} (1998) 550
\smallskip
\item{41.} T.D. Lee and C.S. Wu, Ann. Rev. Nucl. Sci. {\bf 16} (1966) 511;
Errata, {\bf 17} (1967) 513
\smallskip
\item{42.} L. Maiani, $CP$ and $CPT$ violation in neutral kaon decay,
in {\it The Second Da$\,\mit\Phi$ne Physics Handbook},
L. Maiani, G. Pancheri and N. Paver, eds., (INFN, Frascati, 1995)
\smallskip
\item{43.} F. Benatti and R. Floreanini,
Mod. Phys. Lett. {\bf A12} (1997) 1465;
Banach Center Publications, {\bf 43} (1998) 71;
Comment on ``Searching for evolutions
of pure states into mixed states in the two-state system $K$-$\overline{K}$'',
{\tt hep-ph/9806450}
\smallskip
\item{44.} J. Ellis, J.L. Lopez, N.E. Mavromatos and D.V. Nanopoulos,
Phys. Rev. D {\bf 53} (1996) 3846
\smallskip
\item{45.} P. Huet and M.E. Peskin, Nucl. Phys. {\bf B434} (1995) 3
\smallskip
\item{46.} N.E. Mavromatos and R.J. Szabo, Non-linear Schr\"odinger
dynamics of matrix D-branes, {\tt hep-th/9909129}
\smallskip
\item{47.} M. Czachor and M. Kuna, Phys. Rev. A {\bf 58} (1998) 128
\smallskip
\item{48.} F. Benatti and R. Floreanini, Phys. Lett. {\bf B468} (1999) 287
\smallskip
\item{49.} F. Benatti and R. Floreanini, Phys. Lett. {\bf B401} (1997) 337
\smallskip
\item{50.} The CPLEAR Collaboration, Phys. Lett. {\bf B444} (1998) 43;
Eur. Phys. J. C{\bf 5} (1998) 389; Phys. Lett. {\bf B458} (1999) 545
\smallskip
\item{51.} G. D'Ambrosio, G. Isidori and A. Pugliese, $CP$ and $CPT$
measurements at Da$\Phi$ne, in
{\it The Second Da$\,\mit\Phi$ne Physics Handbook},
L. Maiani, G. Pancheri and N. Paver, eds., (INFN, Frascati, 1995)
\smallskip
\item{52.} F. Benatti and R. Floreanini, Mod. Phys. Lett. {\bf A14}
(1999) 1519
\smallskip
\item{53.} The KTeV Collaboration, Phys. Rev. Lett. {\bf 88} (1999) 22
\smallskip
\item{54.} The NA48 Collaboration, Phys. Lett. {\bf B465} (1999) 335

\bye